\documentclass{aa}
\usepackage{graphicx}
\usepackage{txfonts}
\usepackage{natbib}
\bibpunct{(}{)}{;}{a}{}{,}
\newcommand{\Msun}{\mbox{$\rm M_{\odot}$}}
\newcommand{\Msunyr}{\mbox{$\rm M_{\odot}$\,yr$^{-1}$}}
\newcommand{\Lsun}{\mbox{$\rm L_{\odot}$}}
\newcommand{\Rsun}{\mbox{$\rm R_{\odot}$}}

\newcommand{\micron}{$\mu$m}

\def\mathstacksym#1#2#3#4#5{\def#1{\mathrel{\hbox to 0pt{\lower#5\hbox{#3}\hss} \raise #4\hbox{#2}}}}
\mathstacksym\gta{$>$}{$\sim$}{1.5pt}{3.5pt} 
\mathstacksym\lta{$<$}{$\sim$}{1.5pt}{3.5pt} 
\begin{document}
                                %
\title{The origin of mid-infrared emission in massive young stellar
  objects: multi-baseline VLTI observations of W33A\thanks{Based on observations with the VLTI, proposal 381.C-0602}}
\titlerunning{MIDI observations of W33A}
\author{W.J. de Wit, M.G. Hoare, R.D. Oudmaijer, S.L. Lumsden} 
\offprints{W.J. de Wit, \email{w.j.m.dewit@leeds.ac.uk}}
\institute{School of Physics \& Astronomy, University of Leeds, Woodhouse Lane, Leeds LS2 9JT, UK}

\date{Received date; accepted date}
\abstract
{}
{In this paper we aim to determine the structure on 100\,AU scales of the
massive young stellar object W33A, using interferometric observations in the
mid-infrared. This emission could be due to a variety of elements,
for example the inner protostellar envelope, outflow cavity walls, dusty or
gaseous accretion disk.}
{We use the Unit Telescopes of the VLT Interferometer in conjunction with the
MIDI instrument to obtain spectrally dispersed visibilities in the $N$-band on
4 baselines with an angular resolution between 25 and 60 milli-arcseconds
(equivalent to 95 and 228\,AU at 3.8\,kpc). The visibility spectra and
spectral energy distribution are compared to 2D-axi-symmetric dust radiative
transfer models with a geometry including a rotationally flattened envelope 
and outflow cavities. We assume an O\,7.5 ZAMS star as the central source, consistent with the observed
bolometric luminosity. The observations are compared to
models with and without (dusty and gaseous) accretion disks.} 
{The visibilities are between 5\% and 15\%, and the non-spherically 
symmetric emitting structure has a typical size of 100\,AU. A
satisfactory model is constructed which reproduces the visibility spectra
for each (u,v) point. It fits the $N$-band flux spectrum, the mid-infrared
slope, the far-infrared peak, and the (sub)mm regime of the SED. It produces
a 350\,\micron~morphology consistent with the observations.}
{{The mid-infrared emission of W33A on 100\,AU scales is dominated by
the irradiated walls of the cavity sculpted by the outflow. The
protostellar envelope has an equivalent mass infall rate of
$7.5\,10^{-4}\,\Msunyr$, and an outflow opening angle of
$2\theta=20\degr$.  The visibilities rule out the presence of dust
disks with total (gas and dust) masses more than $\rm
0.01\,M_{\odot}$.  Within the model, this implies a disk
$\dot{M}_{\rm acc}$ of less than $1.1\,10^{-7}
(\alpha/0.01)\,\Msunyr$, where $\alpha$ is the viscosity of the Shakura-Sunyaev prescription. However, optically
thick accretion disks, interior to the dust sublimation radius, are allowed to accrete at
rates equalling the envelope's mass infall rate (up to
$10^{-3}\,\Msunyr$) without substantially affecting the visibilities
due to the extinction by the extremely massive envelope of W33A.}}

\maketitle
\section{Introduction}
Circumstellar accretion disks are an essential element in the formation of
stars. The relatively long time scales involved in {\it low-mass} star
formation (SF) and the proximity of the regions where they form allow the
disks to be imaged at millimeter (Dutrey et
al. 1996)\nocite{1996A&A...309..493D}, near-infrared (near-IR; Padgett et
al. 1999\nocite{1999AJ....117.1490P}) or even optical wavelengths (Grady et
al. 1999\nocite{1999ApJ...523L.151G}) and studied in great detail. The short
time scales on which high-mass stars are formed and the distance of massive SF
regions amongst others render the study of the accretion process particularly
difficult. Whether or not a high-mass star grows in mass through the accretion
of matter delivered to the stellar surface by means of a circumstellar disk
is therefore still left unanswered. Recent 3-D radiation hydrodynamic simulations
demonstrate that accretion through a disk can continue in the presence of 
strong radiation pressure (Krumholz et
al. 2009\nocite{2009Sci...323..754K}). Observationally a growing number of
direct and indirect pieces of evidence point to the presence of large-scale
(10,000\,AU) rotating toroids (e.g. Beltr\'{a}n et
al. 2004\nocite{2004ApJ...601L.187B}, 2005\nocite{2005A&A...435..901B}; Furuya
et al. 2008\nocite{2008ApJ...673..363F}), but also to intriguingly disk-like
structures on scales of several 100\,AU in a small number of comparatively
nearby examples (e.g. Hoare 2006\nocite{2006ApJ...649..856H}; Jim\'{e}nez-Serra et
al. 2007\nocite{2007ApJ...661L.187J}; for a review see Cesaroni et al. 2007\nocite{2007prpl.conf..197C}).

The reality of a disk accretion scenario in massive star formation can be
assessed by observations of massive stars during the main mass gathering phase
at significant angular resolution. Prime examples of such young massive stars
constitutes the class of object referred to as massive YSO (MYSO), but also as
high-mass protostellar object, and BN-object. The class is characterised by
luminous ($>10^{4}\,\Lsun$) infrared objects with spectral energy
distributions that peak around 100\,\micron. The luminosity indicates a single
ZAMS star that would be able to ionise its surroundings, however too little or
no radio emission is observed. This absence has been used as an argument to
claim on-going mass infall onto the central object. As a result the star is
extended and thus relatively cool (Hosokawa \& Omukai
2009)\nocite{2009ApJ...691..823H}, and therefore unable to ionise its
surroundings (Hoare \& Franco 2007\nocite{2007dmsf.book...61H}).   
Alternatively, high mass infall rates (Walmsley 1995)\nocite{1995RMxAC...1..137W} 
or gravitational entrapment (Keto 2002\nocite{2002ApJ...580..980K}) could quench the development of an \ion{H}{ii} 
that dominates the radio continuum emission. These alternative scenarios invoke however very high emission measure (EM) \ion{H}{ii} 
regions to make them optically thick in the radio, e.g. $\rm \tau_{1\,cm}= 3\,10^{-10} \times EM\,(cm^{-6}\,pc)$ (Osterbrock 1989\nocite{1989agna.book.....O}). 
However, it is unlikely in this picture
that the region would still be optically thick in the near-IR \ion{H}{i} 
recombination lines Br$\alpha$ and Br$\gamma$ as these have 
line centre optical depths that are $10^{5-6}$ times smaller than $\rm \tau_{1\,cm}$ 
(Hummer and Storey 1987\nocite{1987MNRAS.224..801H}). This would imply that these
objects should have strong near-IR \ion{H}{i} lines and yet
they only show weak, broad lines indicating a stellar wind origin (Bunn et 
al. 1995). Nonetheless, the MYSO class has 
the potential to test the disk accretion scenario. The actual accretion region is probably found on AU to tens of AU scales near
the central stellar object. This region is however buried by hundreds of
$A_{\rm v}$ of dust, which makes up the extended ($\sim $0.1\,pc) protostellar
envelope (van der Tak et al. 2000). Only
long-wavelength radiation is capable of carrying away information on the
physical processes at play, limiting the attainable spatial resolution with single
dish apertures. Interferometers can overcome the problem of limited resolution
and the near and mid-infrared wavelength regions currently deliver  the highest
angular resolution by means of e.g. the Very Large Telescope Interferometer (VLTI).

The inner parts of the dusty MYSO protostellar envelope is heated to a few 100\,K and emits strongly at mid-IR
wavelengths. The 10\,\micron~emission is usually unresolved at the 0.3\arcsec~
resolution of single 8 meter dish telescopes, but it is partially resolved at
24.5\micron~(de Wit et al. 2009).  Here, we present new observations
made of the MYSO \object{W33A} with the MID-infrared Interferometric (MIDI) instrument
at the VLTI. MIDI is a two beam combiner that operates in the $N$-band
(8--13$\mu$m) and delivers spectrally dispersed interferometric observables
(see Leinert et al. 2003\nocite{2003Msngr.112...13L}) . W33A is one of the
most intensely studied MYSO thanks to its IR spectrum rich in ice
  features (Gibb et al. 2000\nocite{2000ApJ...536..347G}). Its physical
nature has received much less attention. The object has an IRAS luminosity of
$10^{5}\,\Lsun$ (Faundez et al. 2004\nocite{2004A&A...426...97F}) for a
kinematical distance of 3.8\,kpc (Bronfman et al. 1996). The object shows
weak, compact and optically thick radio continuum emission (Rengarajan \& Ho
1996\nocite{1996ApJ...465..363R}; van der Tak \& Menten
2005\nocite{2005A&A...437..947V}) and broad ($\sim$100\,km\,s$^{-1}$),
single-peaked \ion{H}{i} recombination emission consistent with an ionised
stellar wind origin (Bunn et al. 1995\nocite{1995MNRAS.272..346B}). Modelling
of the protostellar envelope with spherical dust radiative transfer (RT) model
at constant density by G\"{u}rtler et al. (1991) required the presence of an
inner dust-free cavity of 135\,AU (35\,milli-arcsecond at 3.8\,kpc), a size
that would be well resolved by MIDI. In de Wit et
al. (2007\nocite{2007ApJ...671L.169D}, hereafter paper\,I) we presented a
single MIDI baseline for W33A in order to probe this hundred AU scale
region. We found using 1-D modelling an inner dust-free radius at
7\,milli-arcsecond (26.5\,AU), corresponding to the dust sublimation
radius. We concluded that the dispersed MIDI visibilities and SED can be
reproduced by a spherical dust model provided that the density law is
relatively flat. In this paper we present three additional VLTI/MIDI
observations of W33A with baseline position angles close to perpendicular and
with different projected lengths. This allows us to constrain the distribution
of the emitting material further using 2-D axi-symmetric RT models.

We describe and discuss the new and archived MIDI observations in Sects.\,\ref{Observations} 
and \ref{descrip}. The basic features of the RT model, our modelling approach and 
the final preferred model are explained in Sect.\,\ref{modelling}. The implications for W33A and their effect for
MYSOs in general are presented in Sect.\,\ref{discussion}. We conclude our work in Sect.\,\ref{conclusions}


\section{Observations and data reduction}
\label{Observations}

\begin{table}
  {
     \begin{center}
      \caption[]{MIDI observations of W33A executed in high-sens mode at a
        spectral resolution of $\rm R=30$. Seeing is the one measured by the DIMM. Configuration D was already presented in
        de Wit et al. (2007). }
      \label{tabobs}
      \begin{tabular}{cccccc}
        \hline
        \hline
        &  UT Date    & Stations     & B      & P.A.    & Seeing \\
        &             &              & (m)    & (\degr) & (\arcsec)   \\
        \hline
        Config.A &  2008-06-21 & U2-U3        &  42.03 &  15.04    & 0.65\\
        Config.B &  2008-04-22 & U3-U4        &  57.95 & 119.69    & 1.08\\
        Config.C &  2008-04-21 & U3-U4        &  61.08 & 114.57    & 0.52\\
        Config.D &  2005-09-16 & U2-U3        &  45.58 &  47.31    & 1.00\\
        \hline
        \hline
      \end{tabular}
    \end{center}
  }

\end{table}

\subsection{MIDI observations}
\begin{figure}
  \includegraphics[height=9cm,width=9cm]{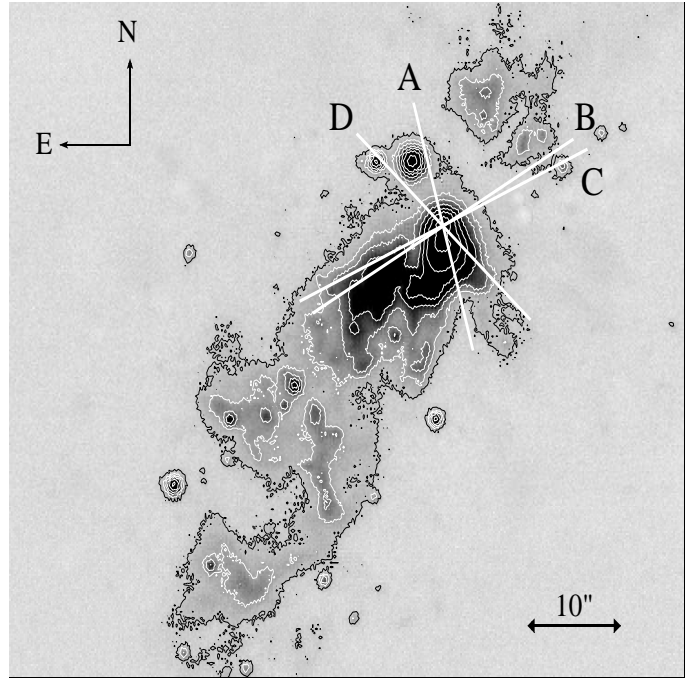}
  \caption[]{UKIDSS $K$-band observations of W33A. Projected VLTI/MIDI 
    baselines are indicated by white bars.}
  \label{kband}
\end{figure}

W33A was observed with the VLTI and the MIDI instrument in service mode on
four different occasions using the UTs as aperture stations (see
Table\,\ref{tabobs}).  We aimed at covering different position angles on the
shortest possible UT baselines. The baselines are presented in
Fig.\,\ref{kband} projected onto the near-IR reflection nebula of W33A as
imaged by UKIDSS (Casali et al. 2007\nocite{2007A&A...467..777C}; Warren et al
2007\nocite{2007MNRAS.375..213W}; Lucas et
al. 2008\nocite{2008MNRAS.391..136L}).  The shortest VLTI baselines are
attained with the ATs but W33A cannot be observed in this mode due to its
relatively low brightness. The obtained angular resolution for the data set
presented here stretches from 25 to 60 milli-arcseconds. Detailed descriptions
of the MIDI observation procedure are given in Przygodda et
al. (2003\nocite{2003Ap&SS.286...85P}), Chesneau et
al. (2005\nocite{2005A&A...435..563C}) and Leinert et
al. (2004\nocite{2004A&A...423..537L}). The W33A observations were executed in
the so-called {\it High-Sens} MIDI mode, which uses all the incoming light for
beam combination and fringe tracking. The incoming beams are interfered
producing two complementary interferometric channels that have by definition a
phase difference of $\pi$ radians. A measurement of the uncorrelated flux
spectrum to obtain final visibilities is done immediately after the
interferometric observation. The high-sens mode is advantageous when observing
faint targets or small visibilities, however the accuracy of the final
visibility spectrum is limited by the sky brightness variation between the
interferometric and photometric measurement. This can amount to 10-15\%
uncertainty, depending on the atmospheric stability. During the nights, the
unresolved interferometric calibrator star \object{HD\,169916} was observed in order 
to determine the instrumental visibility and correct for it. We also inspected
all other observed calibrators from different programs for each night (a total
of 18 objects), which allowed us to estimate the stability and accuracy of the
instrumental visibility. We found it to be uncertain by not more than 5\%, much smaller than the
uncertainty in the determined photometric measurement (20\%, see next
section).  The calibrator star provides also an approximate flux calibration (Cohen et
al. 1999\nocite{1999AJ....117.1864C}) in addition to the visibility
calibration. A prism with a spectral resolution of 30 was employed in order to
disperse the incoming beams.

The correlated flux spectra are estimated by the MIA$+$EWS software package (version
1.6; see Jaffe 2004\nocite{2004SPIE.5491..715J}; K\"{o}hler 2005\nocite{2005AN....326Q.563K}). Interferograms are
coherently added in order to maximize the signal to noise. First, fringe spectra
have to be corrected, because they are affected by the instantaneous atmospheric 
and instrumental piston. The fringe scan is used to estimate
the group delay due to the atmosphere. Removing the atmospheric and the
(known) instrumental group delays constitutes a correction to the
dispersed fringe signal, and straightens the dispersed fringe spectra,
i.e. the phase is independent of wavelength. Next, the phase offset due to
varying water refraction index between the time of recording of the fringe
spectra has to be accounted for. In principle all spectra can then be added to
a final fringe spectrum and the amplitude of the power at each frequency
estimated, which will give the correlated flux.

The total flux spectra are taken in a chopping mode and
each observation results in two spectra as MIDI splits the beams. 
The spectra are extracted from the raw data in the EWS mode. An estimate 
of the sky background is made and subtracted. The area corresponding to 
the mask with which the correlated spectra are extracted is used here.
The counts are then simply summed in the y-direction. The final
spectrum  is the sum of the two geometric means of the four individual spectra:
$\sqrt{A1 \times B1}+\sqrt{A2 \times B2}$. This quantity is also what is obtained for the
correlated flux after beam combination and thus ensures consistency in 
deriving visibilities. 

\begin{figure}
  \includegraphics[height=9cm,width=9cm,angle=90]{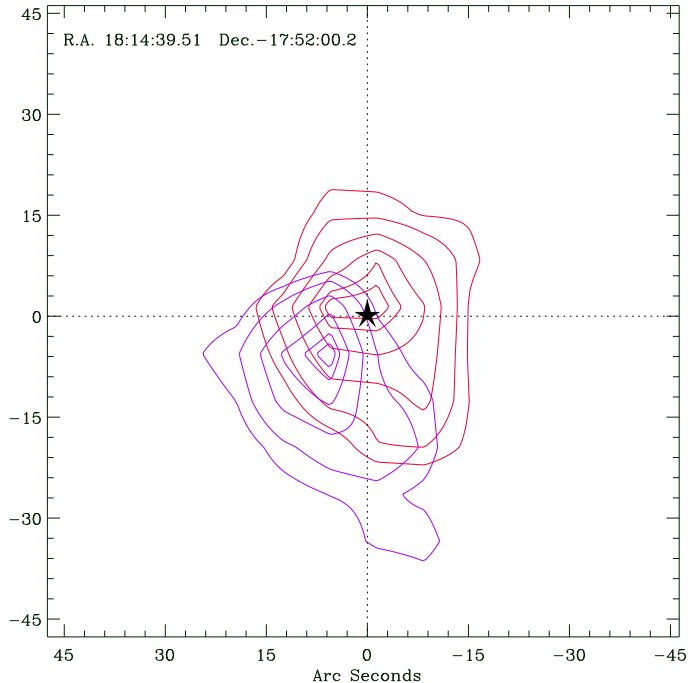}
  \caption[]{$\rm ^{12}CO\,(3-2)$ map taken with JCMT/HARP illustrating the 
molecular outflow of W33A. The P.A. and orientation
of the blue and red-shifted lobes of the outflow are consistent with the $K$-band reflection nebula
 of Fig.\,\ref{kband}. Coordinates are in J2000 and they correspond to the
 UKIDSS near-IR source, and are in accord with the Capps et al. (1978) main IR source.}
  \label{jcmt}
\end{figure}

\begin{figure*}
  \includegraphics[height=18cm,width=14cm,angle=90]{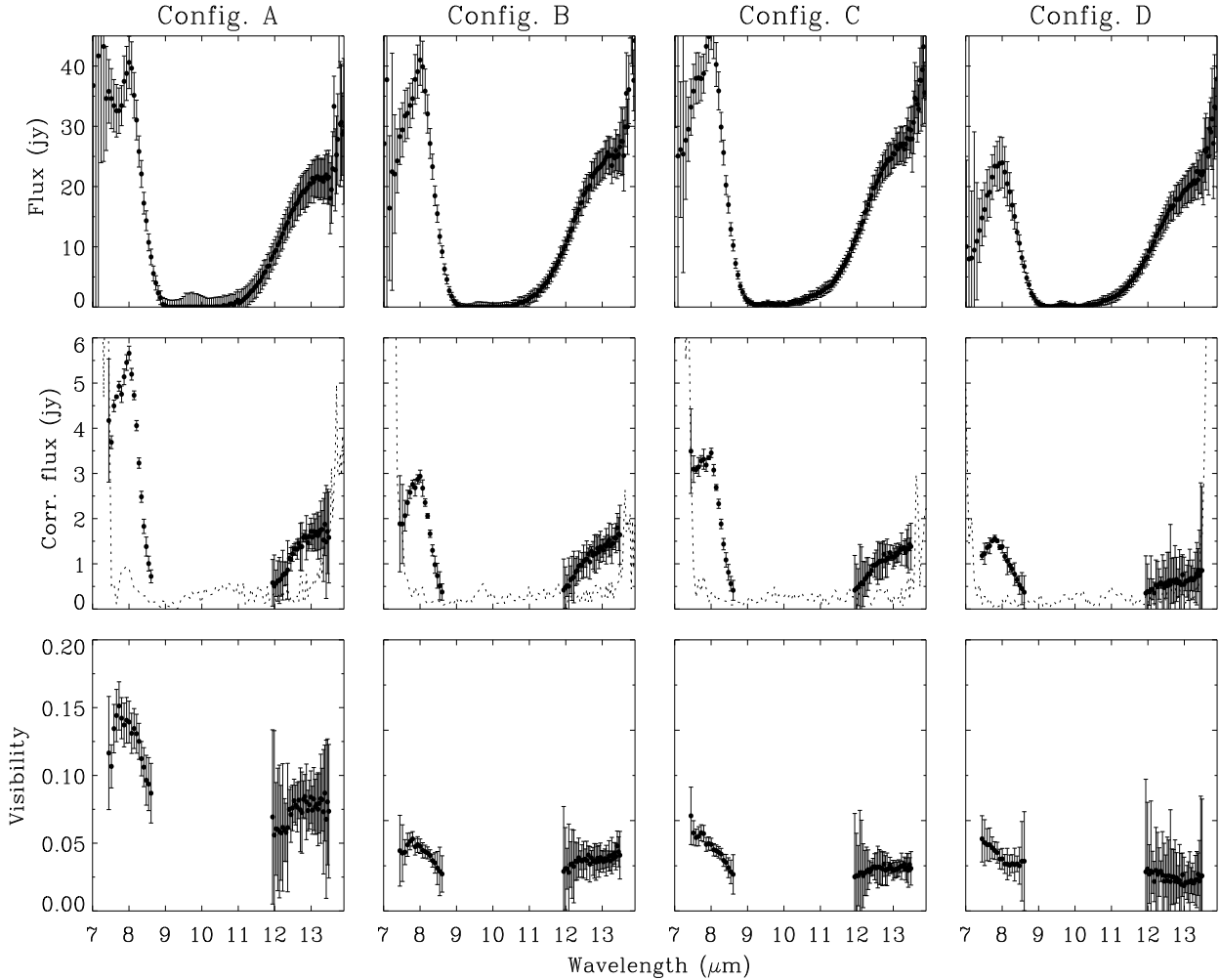}
  \caption[]{MIDI observations of W33A on four different occasions (see Table\,\ref{tabobs}). {\it Top: } 
    Flux spectra. {\it Middle:} Correlated flux spectra (points with errorbars), and the detector noise level (dashed curve). 
   {\it Bottom:} Derived visibilities.}
  \label{spec}
\end{figure*}

\subsection{JCMT/HARP-B observations}
W33A was observed on the night of 5 June 2007 in ${}^{12}$CO, ${}^{13}$CO and
C$^{18}$O using the HARP-B instrument on the James Clerk Maxwell Telescope.  The
${}^{12}$CO data have a resolution of 488\,kHz, and the other two lines were rebinned
to the same resolution, giving a velocity resolution of approximately 0.4\,km\,s$^{-1}$.
The maps had a linear baseline removed.  The core emission was presumed to be
traced by the C$^{18}$O emission (which has line profile very close to a Gaussian).
We used this as a model to trace the emission in the ${}^{12}$CO and ${}^{13}$CO
maps that arises from the actual outflow.  Full details of this work, which is part of a
larger project studying outflows from the Red MSX Source Survey (see e.g. Urquhart
et al. 2008\nocite{2008ASPC..387..381U}), will be published separately in due course.

\section{Description of observational data of W33A}
\label{descrip}
\subsection{$N$-band from MIDI}
The measured MIDI observables (flux, correlated flux) and derived visibilities
of W33A are presented in Fig.\,\ref{spec}.  Each column represents a VLTI configuration and
we have included the observations already presented in paper\,I (configuration
D).  The top panels show the flux calibrated $N$-band spectra. They show a
variation of around 20\% as expected for data taken in the observational
high-sens mode. The errorbars represent two uncertainties that are added in
quadrature: (1) the rms estimates of the
flux and (2) the systematic error determined from a total of five W33A MIDI
spectra (more than the standard one flux spectrum per interferometric
  measurement were taken.). The distinct feature of W33A in the $N$-band is the extremely deep
silicate absorption, the actual depth is not detected at the central
wavelength where it drops to less than 1\% times the level of the pseudo-continuum.

The second row of panels shows the correlated flux spectra.  The dotted lines
indicate the correlated noise level. It was estimated by processing
that part of the detector directly adjacent to the W33A fringes, indicative of
the background noise. The noise produces a signal a few times
0.1\,Jy around 10\,\micron, despite the absence of incoming flux at the base
of the silicate absorption. Similar noise levels have been found previously
(e.g. Jaffe et al. 2004; Matsuura et
al. 2006\nocite{2006ApJ...646L.123M}). The errorbars of the W33A correlated
flux spectra reflect the uncertainties due to spurious correlated noise level.
We discard any signal below this noise level in further analysis. 

The third row of figures presents the resulting visibilities for the wavelengths
not affected by low flux levels. The errorbars are the uncertainties from
correlated and total flux added in quadrature. This entails that the
systematic uncertainty inherent to the high-sens mode are taken into account
in the visibility uncertainties. We find that W33A has visibility values 
between 5\% and 15\% on the used baselines. The spectral shape of the
visibilities reveal a clear trend of decreasing visibilities on the blue 
wing of the silicate absorption feature. The smaller visibilities imply a
 larger emitting region, and therefore that the increasing optical
depth due to the silicate absorption is due to structures on larger size scales than 
the pseudo-continuum. The visibilities corresponding to the two
similar configurations (B and C) are practically the same, as are their
correlated flux levels. It gives an independent indication of the
stability of the set-up, and the relatively good weather. Configuration A
shows a correlated flux level twice higher than the ones of configuration B and C, despite the relatively little
difference in total flux. Given that the weather during the observations was
stable, we believe that the difference in visibilities between the various
configurations is real. If we
translate the visibilities at 8\,\micron~to Gaussian emission regions then
the FWHM decreases from 115\,AU to about 95\,AU rotating in P.A.
from 15\degr (configuration A) to $\sim 120\degr$ (B and C) and changing the
baseline length accordingly.

\begin{figure}
  \includegraphics[height=12cm,width=9cm]{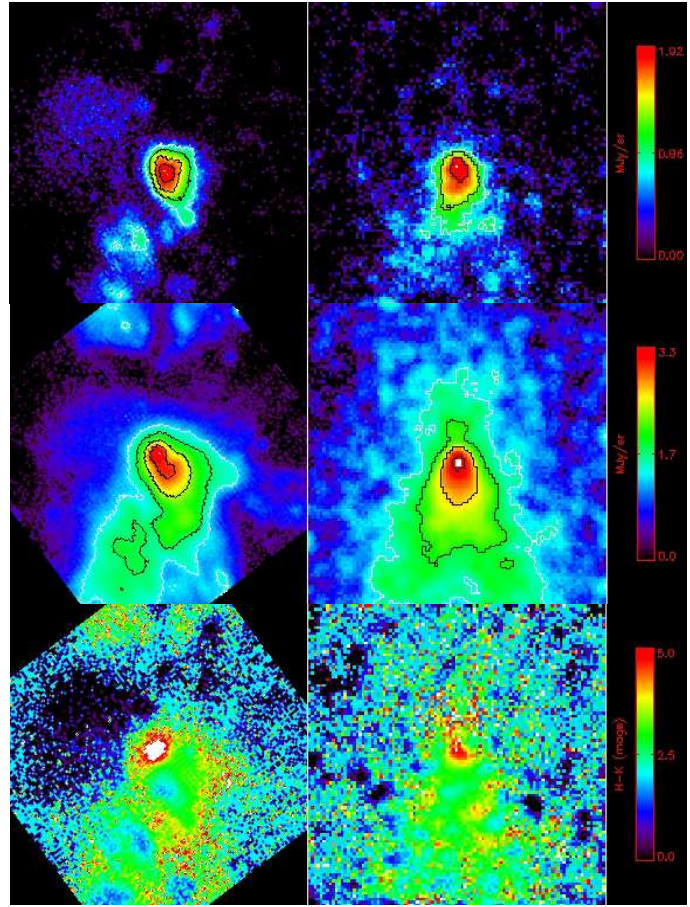}
  \caption[]{{\it Left:} UKIDSS $H$-band (upper panel) and $K$-band (middle panel)
observations of the base of the W33A outflow scattering nebula. Images are
logarithmically scaled, in units of MJy/sr and rotated by $35\degr$ so that the main lobe is pointing
downwards. $H$-band contours are at 7.5\%, 15\%, 40\% and 70\% of peak flux; the
$K$-band contours are at 1\%, 2.5\%, 10\%, and 40\% of peak flux. Secondary stellar
sources have been removed from the images. The lower panel shows the $H-K$ colour images (in magnitudes).
{\it Right:} Model images produced with the RT scattering
code, see Sect.\,\ref{scat}. The model images are convolved
with 2-D Gaussian function with a FWHM of 1.6\arcsec. All images cover an area of $25\arcsec$ on the side.}
  \label{scatmod}
\end{figure}

\subsection{The molecular outflow and near-IR reflection nebula}
The star formation activity in W33A produces an archetypical bipolar
molecular outflow (Fig.\,\ref{jcmt}). The south-eastern lobe constitutes the approaching,
blue-shifted flow. Detailed millimeter line maps which would provide
constraints on the outflow opening angle and the inclination of the system are
currently unavailable. However, outflow activity produces reflection nebulae
particularly well traced by near-IR scattered light (see e.g. Tamura et
al. 1991)\nocite{1991ApJ...378..611T}. The near-IR image of W33A already presented in
Fig.\,\ref{kband}  is consistent with the CO map in position angle and orientation.
We use this image therefore in order to constrain these system parameters, which will be of further help in modelling the MIDI visibilities and SED later
on. The full extent of the $\sim 50\arcsec$ long $K$-band nebula is shown in
Fig.\,\ref{kband}.  The nebula has a position angle of $\sim 145\degr$,
however on scales of order 1\arcsec~patchy extinction distorts the
morphology. A fainter extension can also be seen in the north-western
direction.  This lends support for the $145\degr$ position angle to be the
actual orientation of the system.  High-angular resolution IFU images
presented in Davies et al. (2010\nocite{Davies2010}) confirms the reflected nature of the
$K$-band emission at the base of the outflow and position angle. The outer regions of the nebula
may have contributions from shocked $\rm H_{2}$ emission as traced by extended
4.5\,\micron~ emission in {\it Spitzer} IRAC images (Cyganowski et
al. 2008\nocite{2008AJ....136.2391C}).

In Davies et al. (2010), it is also shown using spectro-astrometry of the
Br$\gamma$ transition that the near-IR source is at the base of the bipolar
outflow on scales of 200\,$\mu$-arcseconds. The near-IR source corresponds
most likely to the central objects in W33A, indeed as was assumed during the
execution of the MIDI observations. This source shows a $K$-band profile in
the UKIDSS image that is broader than the unresolved stars in the nearby
field. In the $H$-band the scattering nebula is much weaker (see
Fig.\,\ref{scatmod}). Given that photons are more efficiently scattered at
shorter wavelength the relatively dim nature of the $H$-band nebula argues for
strong extinction in the line-of-sight towards W33A. Moreover, the central
source is absent in the $H$-band.

Most of the emission seen at $K$-band is dust scattered; contribution
to the emission by molecular hydrogen gas is small as evidenced by
intermediate resolution spectra, obtained as part of the RMS
survey\footnote{URL: http://www.leeds.ac.uk/RMS}. In the $H$-band a foreground
star is visible projected onto the nebula, located 2.5\arcsec~south and
1.0\arcsec~east of the main $K$-band source. We have re-measured the flux
densities coming from the full extended nebula as seen on the UKIDSS images, after first
removing all the interloping stars from the images and replace the
pixels values using the adjacent regions. We find
that in the $K$-band the nebula is 0.39\,Jy ($K \sim 8.0$) and in the $H$-band
0.025\,Jy ($H \sim 11.5$).

\section{Modelling}
\label{modelling}
In this section we present the simultaneous modelling of the visibility and
spectral energy distribution of W33A. Historically, the SEDs of MYSOs have
been interpreted as hot stars embedded in a spherically symmetric dusty
envelope (e.g. G\"{u}rtler \& Henning 1986\nocite{1986Ap&SS.128..163G}).  In
Paper\,I we adopted the approach of modelling the observables with a 1-D
spherical symmetric radiative transfer code. We concluded that models with shallow
density laws are preferred provided that a relatively low effective
temperature for the central star is adopted. Here we 
construct a fiducial model of a hot star surrounded by a protostellar envelope
with outflow cavities. Our initial model aims to be as simple as possible and we
begin by investigating the imprint on $N$-band visibilities by a protostellar 
envelope that includes outflow cavities. The walls of the cavities are
suggested to be the source of mid-IR emission seen on 1\arcsec~scales in the
nearly edge-on MYSO \object{G35.20-0.74} (De Buizer 2006\nocite{2006ApJ...642L..57D})
and in \object{Ceph\,A HW2} (de Wit et al. 2009).

\subsection{Description of the radiative transfer code and input}
We adopt the 2D-axi-symmetric dust RT model by Whitney et
al. (see for details Whitney et al. 2003a\nocite{2003ApJ...591.1049W}, 2003b\nocite{2003ApJ...598.1079W}). The model
calculates radiation transfer (absorption, re-emission and scattering) through
a dusty structure.  The inner rim of the structure is the dust sublimation
radius which follows from a fit to models with different stellar temperatures
(see Whitney et al. 2004\nocite{2004ApJ...617.1177W}: $R_{\rm subl}/R_{*}=(T_{\rm
subl}/T_{*})^{-2.1}$); space is empty within the inner rim except for the star
itself. The structure
can include three distinct geometrical elements: an accretion disk, a
protostellar envelope, and low density polar outflow cavities. The
protostellar envelope is described by the analytical TSC solution (Ulrich
1976\nocite{1976ApJ...210..377U}; Tereby, Shu \& Cassen 1984\nocite{1984ApJ...286..529T}) of a simultaneously rotating and collapsing spherical
structure. This solution is governed by the infall parameter ($\dot{M}_{\rm
infall}$) and the centrifugal radius $R_{\rm c}$, i.e. the radius where rotational motion
dominates infall. The envelope mass infall rate is a parameterization of the
density distribution and does not per se represent an actual determination of
infalling material.  The presence of an accretion disk is
optional in the code. When present, the disk structure follows the one for a flared
$\alpha$-type disk prescription, and the released accretion luminosity within
the dust inner rim is added to the central luminosity source. The outflow cavity geometry can have two possible shapes, a
streamline or polynomial geometry. The streamline is conical on large scales,
the polynomial has a $180\degr$ opening angle at the stellar surface. Each of
these geometrical elements are potential mid-IR emitters and could contribute
to the visibilities. For most model runs the $R_{\rm subl}$ is comparatively small (tens of AU) and
unresolved by the employed VLTI baselines in the $N$-band.

A model run is necessarily characterised by a large number of free
parameters. A fraction of these can be constrained by 
the known properties of W33A.  The spatial information at milli arcsecond angular
resolution provided by MIDI allows the RT model to be
seriously tested on its correctness in describing a MYSO.

\subsubsection{Fixed input parameters} 
The central star parameters are chosen close to an O7.5 ZAMS according
to the scale  of Martins et al. (2005)\nocite{2005A&A...436.1049M}. The  bolometric
luminosity is $\rm 10^{5}\,L_{\odot}$ (Faundez et al. 2004) and it is
located at the kinematical distance of 3.8\,kpc (Bronfman et
al. 1996\nocite{1996A&AS..115...81B}). The luminosity is based on IRAS flux density, and could be
overestimated. We take the opportunity here to correct our 70\,\micron~{\it Spitzer} MIPS
photometry presented in Paper\,I, which was not corrected for detector
non-linearity. Taking this correction into account we find a value of $1.9\,10^{3}$\,Jy,
quite similar to the IRAS measurement at 60\,\micron. This indicates that the
luminosity based on IRAS is probably reasonable. We think it is unlikely 
that a significant fraction of the IR luminosity is due the secondary
millimetre source noted in van der Tak et al. (2000) at 6\arcsec~from the main
source. The W33A MIPS 24\,\micron~and 70\,\micron~data show no sign of a secondary object, nor does a new 24.5\,\micron~image taken with 
Subaru/COMICS at an angular resolution of 0.6\arcsec~(in prep.). This
suggests that the IR luminosity is dominated by a single object.

\begin{table*}
  {
    \begin{center}
      \caption[]{Parameters for the preferred model discussed in Sect.\,\ref{themod}. This model consists of an envelope with outflow cavities; a disk
is not present.. The inner radius was chosen equal to the dust sublimation radius. The model has a centrifugal radius of 33\,AU, and cavity opening angle of $2\theta=20\degr$. Most parameters are self-explanatory. $\dot{M}_{\rm infall}$ is the envelope's mass infall rate; $i$ is the viewing angle
with respect to the outflow axis; $R_{\rm env}$ is the envelope's outer radius.}
      \label{mod}
      \begin{tabular}{cccccccccccc}
        \hline
        \hline
          $M_{*}$       & $R_{*}$        & $T_{\rm eff}$ & $M_{\rm env}$   & $\dot{M}_{\rm infall}$  &  $i$  & $R_{\rm env}$ &  $R_{\rm sub}$ & $A_{\rm V}^{\rm for}$  & $A_{\rm V}^{\rm env}$  & $n_{\rm out}$ & P.A. \\  
          (\Msun)        & ($\rm \Rsun$)  & (K)           & (\Msun)         &  ($\Msunyr$)       &  ($\degr$) & (AU)     &  (AU)        &  &   &  ($\rm cm^{-3}$)&($\degr$)\\                  
        \hline
         25            &  8.4          & 35\,000      &  $9.7\,10^{3}$           & $7.5\,10^{-4}$           &   60     & $5\,10^{5}$     &   25            &      8      &230           & $1.4\,10^{4}$&165\\    
        \hline
        \hline
      \end{tabular}
    \end{center}
  }
\end{table*}

\begin{figure*}
  \includegraphics[height=9cm,width=7cm,angle=90]{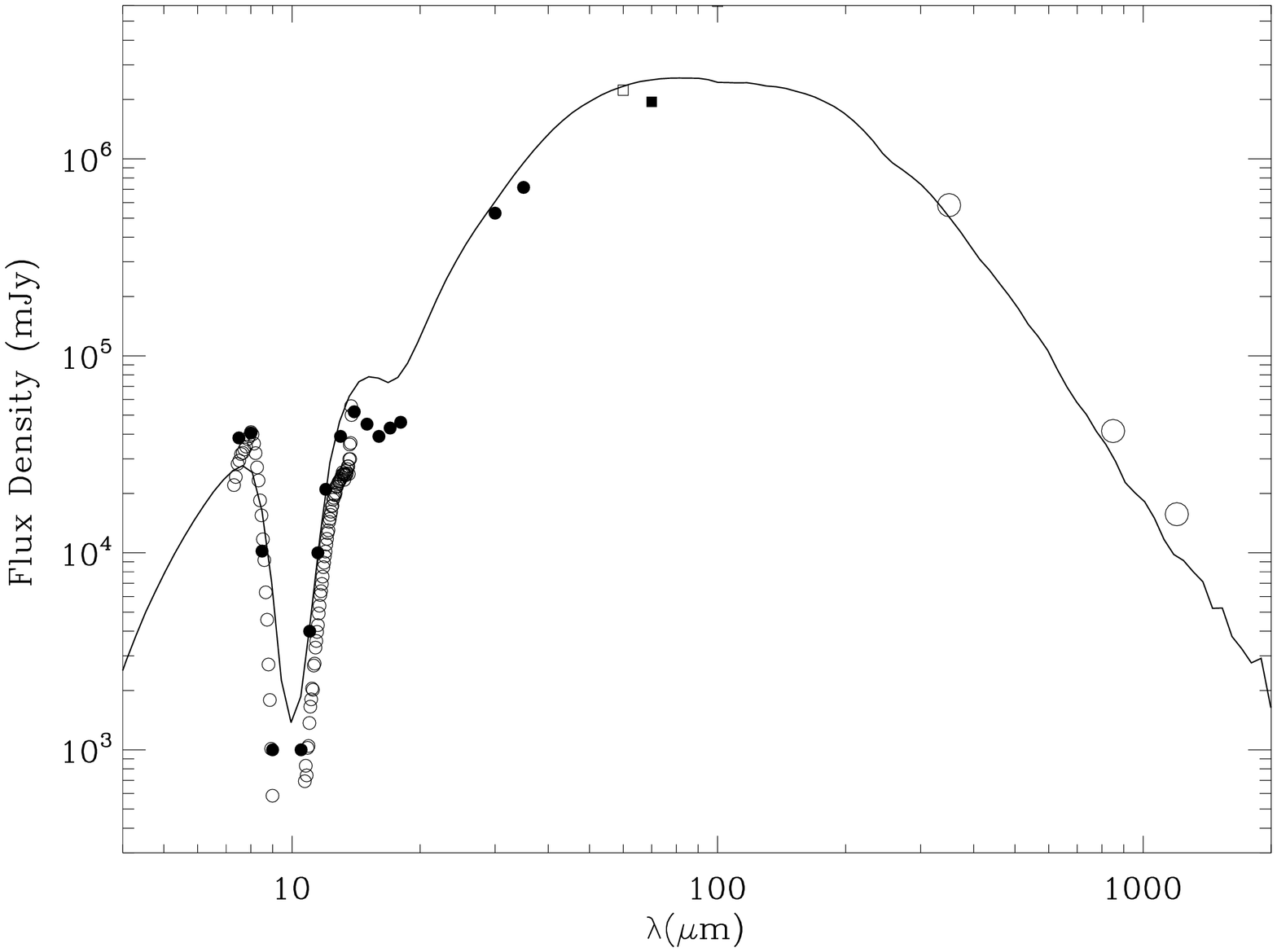}
  \includegraphics[height=9cm,width=7cm,angle=90]{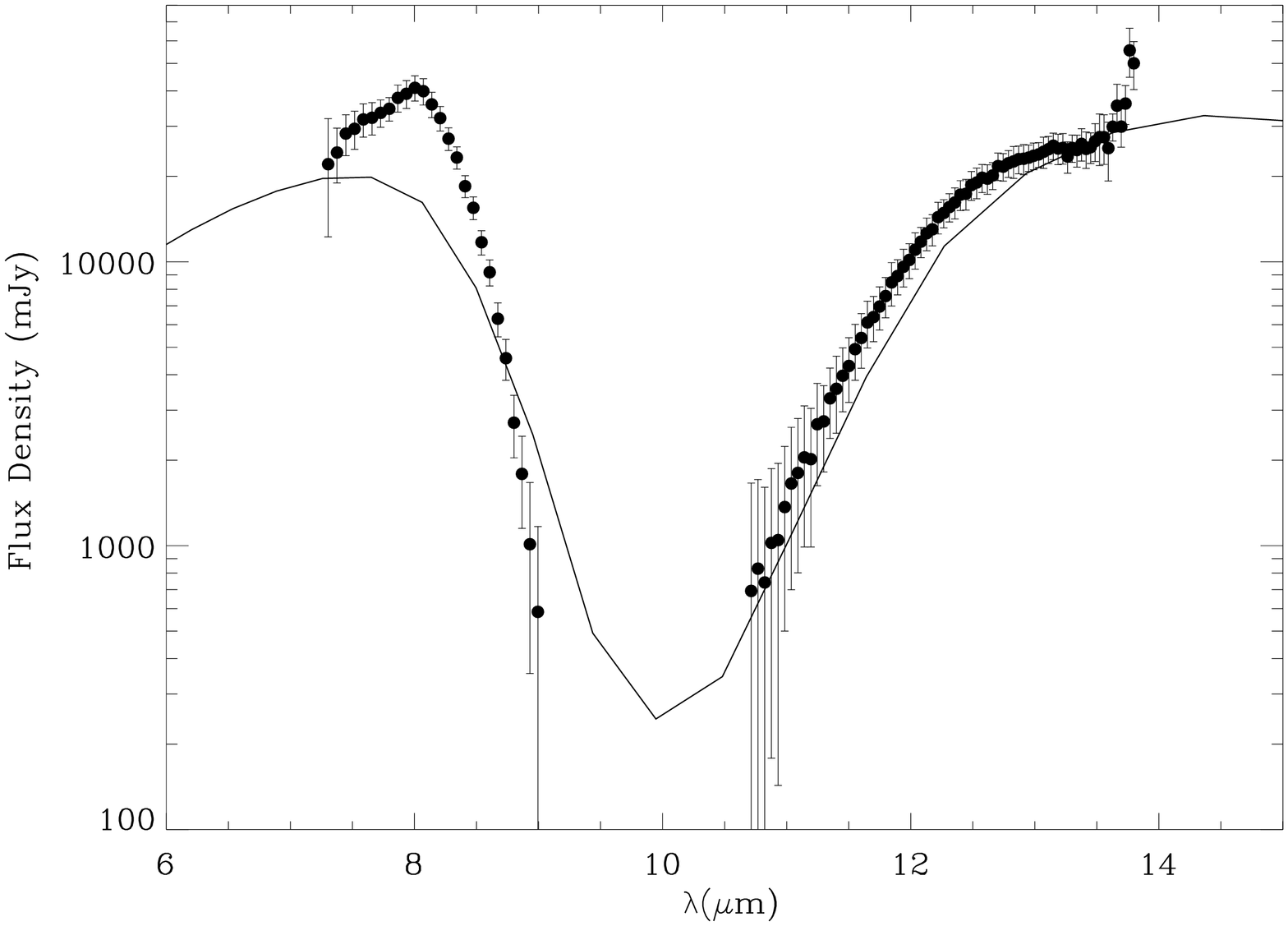}
  \caption[]{{\it Left:} SED of the preferred model overplot W33A flux
densities. Small filled circles trace the ISO SWS spectrum, small open circles are the
MIDI spectrum, filled square is the {\it Spitzer} MIPS point, open square is
the IRAS measurement at 60\micron. In the (sub)mm range, the integrated fluxes at
350\,\micron, 850\,\micron, and 1.2\,mm are from  van der Tak et al (2000), Di Francesco et
al. (2008)\nocite{2008ApJS..175..277D}, and  Fa\'{u}ndez et al. (2004) respectively. {\it Right: } Zoom in onto the $N$
band spectrum. Shown are the MIDI flux spectrum with errorbars and the model fluxes appropriate 
for the MIDI slit width of 0.6\arcsec}.
  \label{fits1}
\end{figure*}

A number of the important parameters that determine the properties of the TSC
envelope are fixed for the following reasons. We choose a single type of
dust that consists of an interstellar mixture of ``warm silicates'' (Ossenkopf
et al. 1992\nocite{1992A&A...261..567O}), and amorphous carbon with a size
distribution according to Mathis et al. (MRN; 1977\nocite{1977ApJ...217..425M}).  The type of 
silicates provides a better fit to the observed shape of the absoprtion feature than
  MRN DL grain-mixture (Draine \& Lee 1984\nocite{1984ApJ...285...89D}), in particular to the blue wing (see paper\,I, and see
  also Capps et al. 1978\nocite{1978ApJ...226..863C}; G\"{u}rtler \& Henning
  1986\nocite{1986Ap&SS.128..163G}).  The centrifugal radius is set at 33\,AU which is comparable
to the dust sublimation radius.  TSC envelopes make a transition from a pure
infall, $-1.5$ radial density law to a shallower rotating, $-0.5$ law at the
centrifugal radius. A TSC envelope with an centrifugal radius smaller than the 
sublimation radius is dominated by the infalling part of the envelope's
structure solution, i.e. the envelope is quite spherically symmetric apart
from the outflow cavities.  The shape of the outflow cavities is polynomial (see Whitney et al. 2003),
which in the presence of a stellar wind could be a more appropriate geometry
than a conical streamline.  The density in the outflow cavity is chosen to
be typical of MYSO outflows, i.e. between $5\,10^{3}$ and a few times $\rm
10^{4}\,cm^{-3}$ (see Beuther et al. 2002\nocite{2002A&A...383..892B}). The cavity material
consists of gas and dust. We initially base the outer edge of the envelope on 350\,\micron~ observations by van der Tak et
al. (2000\nocite{2000ApJ...537..283V}). The radial intensity profile shows a downturn of submm
emission at a radial distance of $\sim50\arcsec$, or $\rm 2\,10^{5}\,AU$. We
require a posteriori that the model densities at the outer edge are similar to
that of the large scale molecular cloud material in which W33A is embedded,
viz.  $\rm 10^{4}\,cm^{-3}$.

\subsubsection{Cavity opening angle} 
\label{scat}
One can exploit the observed shape of the near-IR scattering nebula to
constrain the opening angle of the outflow. To this end we use the
less computationally intensive scattering-only version of the dust RT
code (Whitney \& Hartmann 1992\nocite{1992ApJ...395..529W},
1993\nocite{1993ApJ...402..605W}). The scattering code provides images
at ten inclination angles $\mu$ simultaneously, which one can readily
compare to the observations. We model the near-IR nebula under the
assumption that the emission is free from direct thermal emission. We
put particular emphasis on matching the outflow opening angle with the
observed one and constraining the inclination angle. Model input
parameters are conform the fixed parameters discussed in the previous
subsection, except for the dust properties. The detailed dust properties 
are not important for the goal to constrain the geometrical parameters
of the W33A system from the near-IR nebula. For completeness, the scattering dust used here
is coated with a small layer of water ice. The dust sizes follow a
distribution that fits the extinction curve from Cardelli et
al. (1989\nocite{1989ApJ...345..245C}) for a prescription of $R_{\rm
V}=4.0$. This distribution is not a single power law, and it is
presented in Kim et al. (1994\nocite{1994ApJ...422..164K}). For more
details on the dust properties see Fig.\,1 in Whitney et al. (2003a).

\begin{figure*}
  \includegraphics[height=18cm,width=15cm,angle=90]{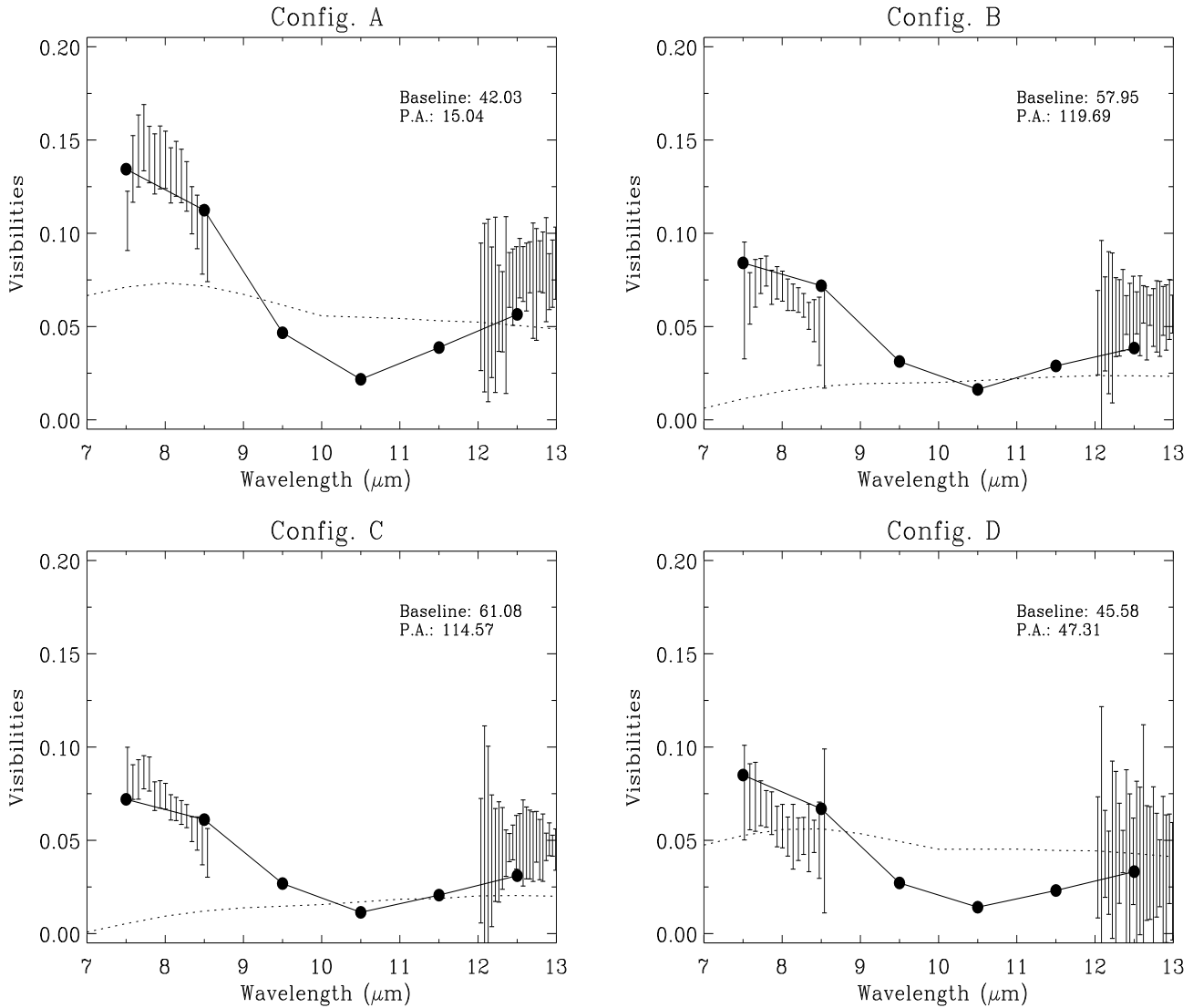}
  \caption[]{Model (full line and filled circles) fits to the observed
visibilities. Model parameters are listed in Table\,2. Corresponding images
are presented in Fig.\,\ref{thmod}. Note the discrepancy with the spherically
symmetric model (dotted lines) used to fit configuration D in paper\,I.}
  \label{visib}
\end{figure*}

Fig.\,\ref{scatmod} compares the $H$ and $K$-band images of a model
run with the W33A scattering nebula. The images of the observed nebula
have been rotated by $35\degr$. The observed and model images are in
the same units (MJy\,sr$^{-1}$). Model images are convolved with a
Gaussian point spread function that simulates the one derived from the
UKIDSS images and Gaussian noise derived from the observed images is
added to the model ones. The outflow angle measured from the images
directly would be $45\degr$, however the presented model has an
envelope geometry with a full cavity opening angle of less than half
that, viz. $2\theta=20\degr$.  In our fitting procedure, attention is
devoted to the absence of an $H$-band counterpart of the central
source. Whether or not a central source is detectable depends on the
system inclination. The model that is presented in Fig.\,\ref{scatmod}
has an inclination of $50\degr$. An inclination less than $40\degr$ or
more than $70\degr$ are excluded. In the former case the central star
is visible in the $H$-band, whereas in the latter case the redshifted
outflow lobe becomes prominent. These features are not present in the
UKIDSS near-IR images.

Although the shape on the sky can be reproduced reasonably well, not a single
model run reproduces the observed $H-K$ colour of $\sim 3.5$. Photometry done 
on the model images shows that they produce colours that are much bluer
instead.  The model images of Fig.\,\ref{scatmod} match the colour
of the observed images, but additional foreground extinction of $A_{\rm
  V}\approx25$ is required. The model $H$ and $K$-band fluxes need to be scaled
up accordingly. The central source visible in the $K$-band constitutes only a minor contribution to the
total observed flux density at this wavelength. We note that MYSO models invariably have problems matching
especially the short wavelength fluxes of MYSO scattering nebulae (e.g. Alvarez et
al. 2004\nocite{2004A&A...419..203A}). The deviant model colours might be
rectified by the inclusion of a clumpy structure within the outflow cavity
(Indebetouw et al. 2006\nocite{2006ApJ...636..362I}) and/or
very small dust particles that are transiently heated to high temperatures. We
find that the introduction of an outflow cavity with a smooth density
distribution into an infalling envelope cannot be the only physical element
determining the observed near-IR emission of MYSOs.  Whichever the solution, a
deeper investigation of this problem is beyond the purpose of the paper.
Nonetheless, a comparison of near-IR images with scattering RT models has led
to relatively stringent constraints on the outflow opening angle, and
comparatively less stringent limits on the inclination.

\subsubsection{Methodology} 
The prime model parameters which remain to be determined are the density as
parametrized by the envelope mass infall rate, the presence of an accretion
disk (and its properties), and the outer radius. They determine the peak of
the SED, and depending on inclination angle, the depth of the silicate
feature, and the dominant emission regions on scales of 100\,AU.  The RT code
is computationally demanding which excludes a grid search for the optimum
value for the decisive model parameters. Building on a simple fiducial model,
we try to find a model that fits the SED and MIDI visibilities
simultaneously. One run delivers the SEDs at 10 inclinations simultaneously,
evenly spaced in $\cos(\mu)$. Model images are produced for a single
inclination at a time but for any number of desired wavelengths/filters. A
total of $4\,10^{7}$ "photons" are used to produce images, for SED-only-runs
the code uses four times less. We divided the $N$-band up in 6 square filters of
1\,\micron~width, running from 7.5\,\micron~to 12.5\,\micron.  The resulting
narrow-band model images are multiplied with a Gaussian model for the VLT-UT
Airy disk. This represents the aperture field of view with a FWHM appropriate
for the particular wavelength. The resulting image is then Fourier transformed
and the visibilities extracted for the projected interferometer baseline and
position angle.  The model images have a pixel size of $0.01\arcsec$ which is
less than half the angular resolution (0.024\arcsec) at the longest baseline
for the shortest wavelength.  The SED is matched from $\sim$8\,\micron~ to
1\,mm. The input data are the MIDI flux spectrum, the slope of the SED
around 40\,\micron~determined from the ISO-SWS spectrum (see Paper\,I) and the
MIPS 70\,\micron~flux density. At the longer wavelengths, the integrated
fluxes from the following observations are used: 350\,\micron~from with the
CSO/SHARC instrument (van der Tak et al. 2000), 850\,\micron~ with JCMT/SCUBA (Di
Francesco et al. 2008), and 1.2\,mmm with SEST/SIMBA (Fa\'{u}ndez et
al. 2004). The far-IR and (sub)mm observations are compared to model SEDs 
representing all the emitted energy at each wavelength, whilst the MIDI flux spectrum is
compared to the model SED with an aperture of 0.6\arcsec~corresponding to the
MIDI slit width. Matching of the SED and visibilities is performed by eye.

\begin{figure*}[t]
  \includegraphics[height=9cm,width=7.5cm,angle=90]{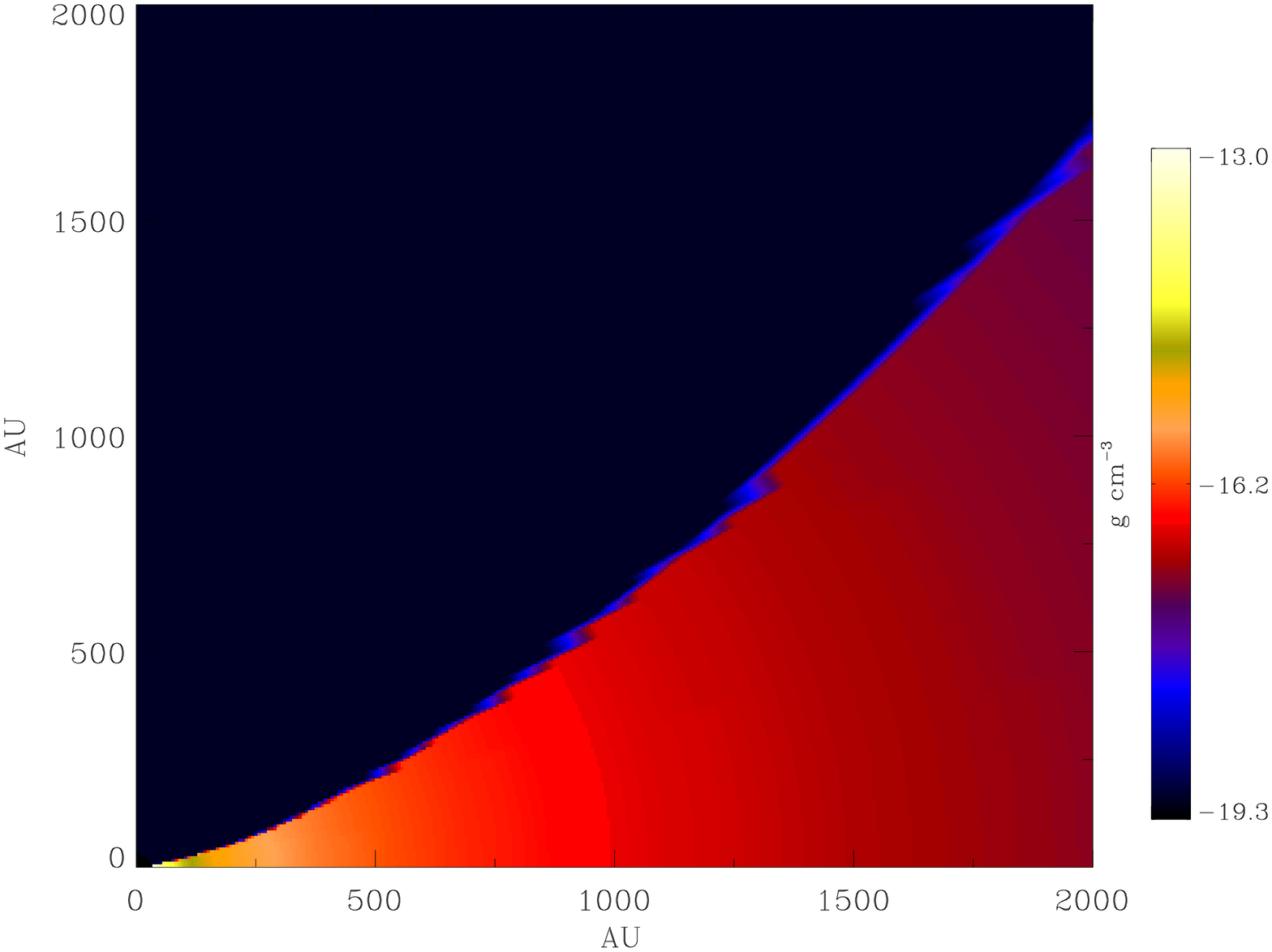}
  \includegraphics[height=9cm,width=7.5cm,angle=90]{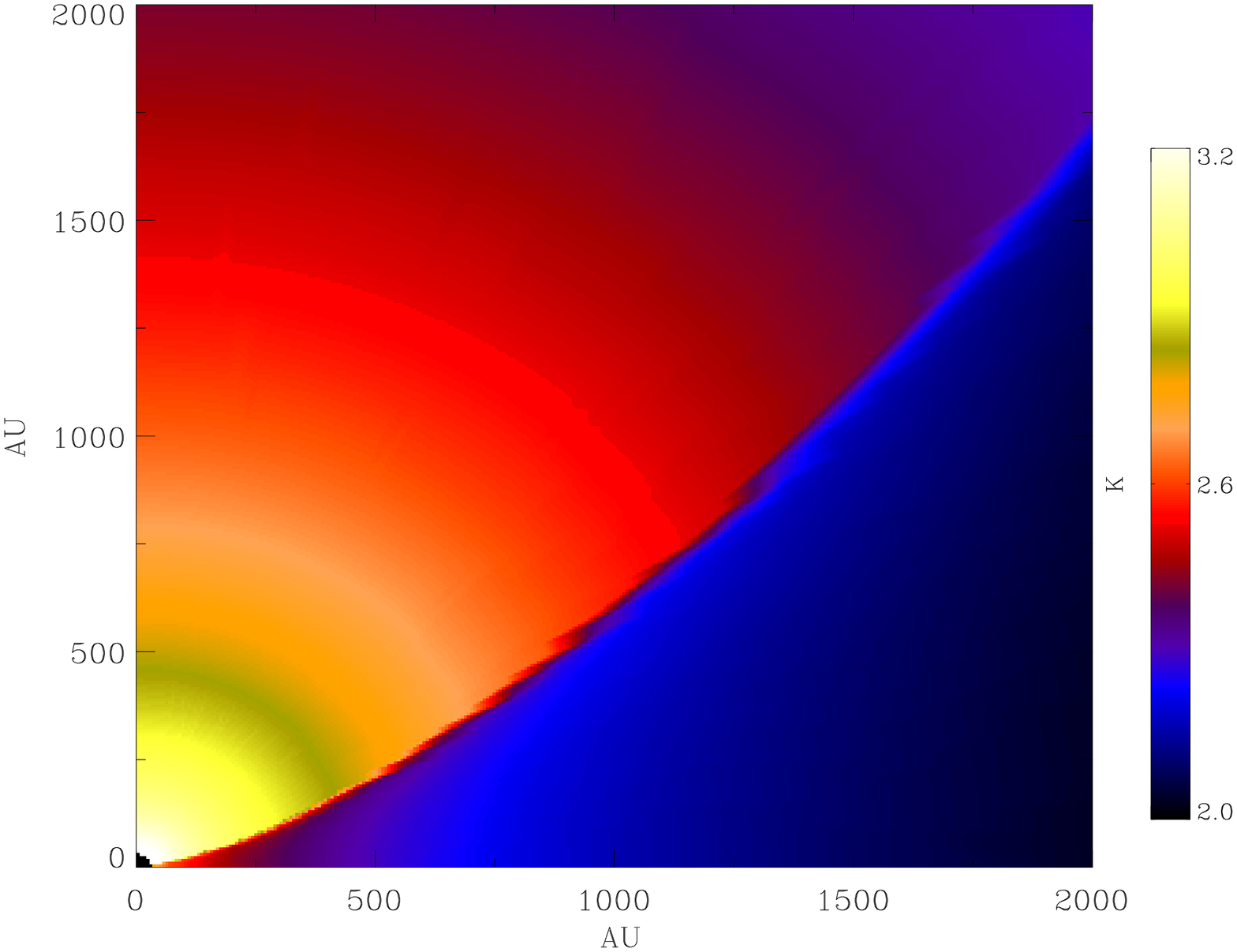}
  \caption[]{Logarithmically scaled slices through the spatial distribution of
    the density (left) and temperature (right) of the preferred model perpendicular
to the outflow axis. The cavity has a very small, but non-zero density consisting of gas and dust (see text).}
  \label{tr}
\end{figure*}

\begin{figure*}[t]
\includegraphics[width=18cm,angle=0]{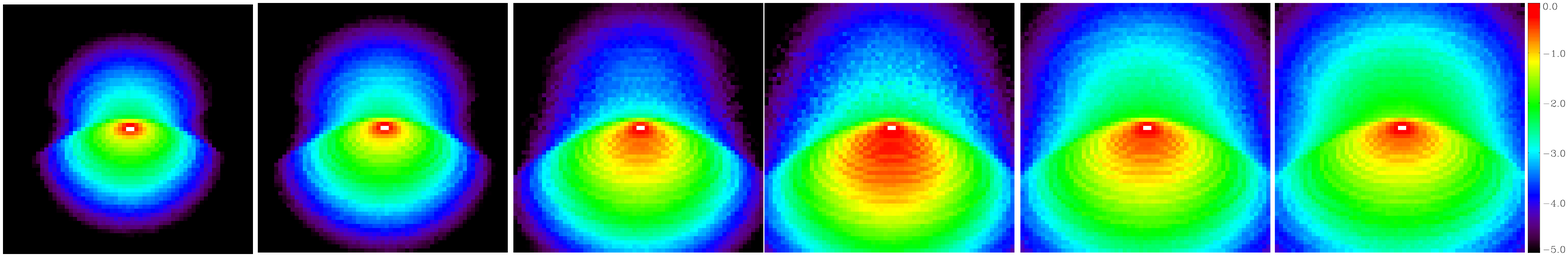}
\includegraphics[width=18cm,angle=0]{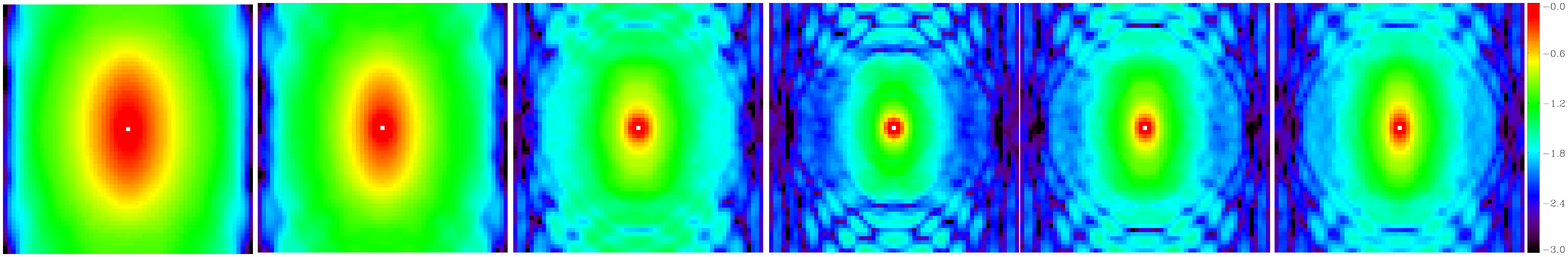}
  \caption[]{Logarithmically scaled images (top row) and corresponding Fourier transforms
    (bottom row) of the preferred model (Table\,\ref{mod}) at wavelengths 7.5\,$\mu$m to 12.5\,$\mu$m in
    steps of 1\,\micron. Image field of view is $0.6\arcsec$ (2280\,AU) on a
    side, corresponding to the MIDI slit width. Pixel scale is 0.01\arcsec. The model 
    images have been multiplied with a wavelength dependent 2-D Gaussian function, simulating the VLT-UT airy disk. Warm
cavity walls dominate the emission at 10\,\micron.}
  \label{thmod}
\end{figure*}

\subsection{Preferred model fit}
\label{themod}
With the parameter values described in the previous section, a model is
obtained that produces a reasonable match to both the SED and the
visibilities. We have run a total of some 30 models in order to narrow down the
parameter values varying the mass infall
rate, inclination angle, and the outer radius. A satisfactory model is presented in Figs.\,\ref{fits1} and
\ref{visib}. The parameter values of this particular model can be found in
Table\,2. The figures show that the model fits both the SED and visibilities
reasonably. A discrepancy between model and observed SED is found in the
near-IR, as expected (see Sect.\,\ref{scat}).  The model reproduces quite well
the spectral trend of smaller visibilities in the wings of the silicate
feature, especially clear on the blue wing. Fig.\,\ref{visib} also displays
the visibilities from the spherically symmetric solution that matches the SED
and visibilities of configuration D only (from Paper\,I). Each panel shows the
spherically symmetric model visibilities calculated for the appropriate
baseline length. It is clear that they provide a considerably worse fit to the
visibilities obtained on the additional configurations A, B and C,
illustrating the need for multiple baselines.

The actual model images of the preferred model are shown in Fig.\,\ref{thmod},
along with a slice through the spatial temperature and density distribution in
Fig.\,\ref{tr}. They make clear that the dominant emission regions in the
$N$-band at a few 100\,AU scale are the envelope regions close to the surface
of the outflow cavity. These envelope regions are warmed up due to irradiation by the star. We refer to these regions as ``cavity walls''
for brevity.  The density distribution in Fig.\,\ref{tr} illustrates that the
density of the paraboloidal cavity is a few order of magnitude less than that
of the envelope. The cavity itself is hotter than the bulk of the envelope but
contributes very little to the $N$-band emission due to the low density.
Fig.\,\ref{thmod} shows that the emission is relatively extended on the 
probed scales and produces consequently very low visibilities. An increase in optical depth, by either probing a
higher inclination angle or increasing the mass infall rate, extinguishes the
warmest, inner part of the irradiated walls. As a result visibilities
generally decrease with an optical depth increase. However the extremely deep
silicate absorption feature of W33A requires a large optical depth. This is
the trade off in order to obtain a decent fit. We find that for an inclination
angle larger than $35\degr$ the mass infall rate needs to be smaller than
$1\,10^{-3}\,\Msunyr$ in order to see the warm cavity wall regions and avoid
spatially resolving the $N$-band emission out.  The increase in optical depth
due to the silicate absorption has a similar effect.  Fig.\,\ref{thmod} makes
clear that the warm inner regions of the cavity walls become progressively
fainter with wavelength causing an increase of the typical emitting size and
hence lower visibilities. 

The same effect is found when one increases the centrifugal radius. The
preferred model with an increased $R_{\rm c}$ to a value of 300\,AU shows
that the density in the equatorial region increases but the increase is
confined to within a fraction of a degree of the equator. The density actually
decreases away from the equator allowing a larger area of the cavity walls to
be heated up. Visibilities drop therefore with increasing $R_{\rm c}$. Further
fine tuning of the model parameters would require additional observations at
preferentially shorter baselines.

\begin{figure}
  \includegraphics[height=8.5cm,width=8cm,angle=90]{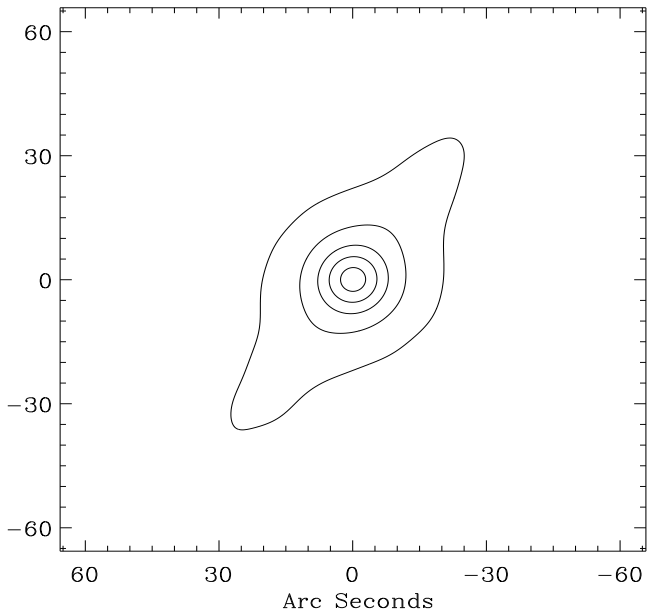}
  \caption[]{Morphology at 350\,$\mu$m predicted by the preferred model. Contours are at 15\%, 30\%,
    50\%, 70\% and 90\% of the peak value. Model image has been convolved with
    a 10\arcsec~beam. See van der Tak et al. (2000) for the observed 350\,$\mu$m image.}
  \label{350mu}
\end{figure}

In summary, MIDI visibilities and SED of W33A can be matched with a simple
geometry that consists of a protostellar envelope with a $-1.5$ radial density
distribution that has polar outflow cavities with a $2\theta=20\degr$ opening angle.
We note that the modelling of the MIDI data is not very sensitive to the exact density
powerlaw in the envelope within the framework of the applied RT model. The
density powerlaw in this case is effectively governed by the centrifugal
radius. Spherical models as applied to W33A in de Wit et al. (2007) are
inadequate in modelling structure on 100 AU scales as Fig.\,\ref{fits1}
illustrates. On the other hand, resolved observations of MYSOs at $\sim 25\,\mu$m at
average inclination angles (de Wit et al. 2009) and at $350\,\mu$m (van der
Tak et al. 2000) provide insight into the density 
law starting at radii of 1000\,AU and indicate a radial density law of
$-1.0$. Matching images of W33A with model images at these respective
wavelengths will provide much stronger constraints on the centrifugal
radius. For now we show in Fig.\,\ref{350mu} the 350\,\micron~morphology of our
model. We have convolved the model image with a 10\arcsec~beam appopriate for
comparison with the CSO observations presented in van der Tak et
al. (2000). The latter observations show a structure more flattened than the
model image, however the FWHM of the model image of 16\arcsec~is bracketed by 
the observed FWHM range of 14.4\arcsec$-$19.4\arcsec, showing a good
correspondance. The density at
the outer radius of the model complies with expected densities of
molecular clouds. Observations
in the submm give a lower limit to the size, while in 
reality the outer parts of the envelope may become even steeper than $-1.5$
(e.g. Mueller et al 2002\nocite{2002ApJS..143..469M}). The outer radius is
therefore uncertain by a factor of 2. The best matching RT model presented does not need 
contributions to the emission from other components within the MYSO environment, instead
$N$-band emission on 100\,AU scale is completely dominated by the warm cavity
walls.

\section{Discussion}
\label{discussion}
In several studies the accretion based formation scenario for massive stars
has been forwarded partially based on observations with 1\arcsec~angular
resolution and/or SED studies of MYSOs (e.g. De Buizer et
al. 2005\nocite{2005ApJ...635..452D}). If MYSOs are to be considered young massive
stars still in their accretion phase, and this accretion occurs in a similar
fashion as in low-mass stars, then circumstellar accretion disks ought to be
present. The MYSO accretion disk is expected to be luminous (approaching that
of the star itself), and estimated to have a size a few times 100\,AU. Our presented 
analysis prompts the question to what degree an accretion disk can contribute
to the mid-IR emission.

\subsection{The contribution of an accretion disk}
The dust RT code allows the inclusion of a dust disk truncated at the 25\,AU
dust sublimation radius. As a first step we consider the effect on the model
fit by adding such a dusty disk to the preferred model presented previously
(Table\,2). We emphasize that the disk is directly irradiated by the star, as
the space between disk and star is empty in the model. The mass of the dusty
disk is initially chosen to be 1\% of the stellar mass, i.e. 0.25\,\Msun. 
  The radial density law of the disk concurs with that expected of an $\alpha$-type disk (i.e. a
power law exponent of 1.875, Frank et al. 1985\nocite{1985apa..book.....F}).
Due to the large inner truncation radius, the actual accretion luminosity
generated by the dust disk is marginal and on the order of 1\% of the stellar
luminosity (see eq. 6 in Whitney et al. 2003a). The formal accretion
rate (see eq. 5 in Whitney et al. 2003a) for $\alpha=0.01$ is then
$2.8\,10^{-6}\,(\alpha/0.01)\,\Msunyr$. We note that the accretion luminosity inside the dust
sublimation radius, where no actual disk structure is present in the RT code,
is emitted with the stellar spectrum (see Whitney et al. 2003a). The disk has
a constant opening angle, i.e. the scaleheight increases linearly with radius.
The model disk outer radius is arbitrary because it does not emit in the
$N$-band, and is better constrained  by millimetre data; for now, it is
set to 500\,AU.
%
%
%
%

\begin{figure}
  \includegraphics[height=8.5cm,width=7cm,angle=90]{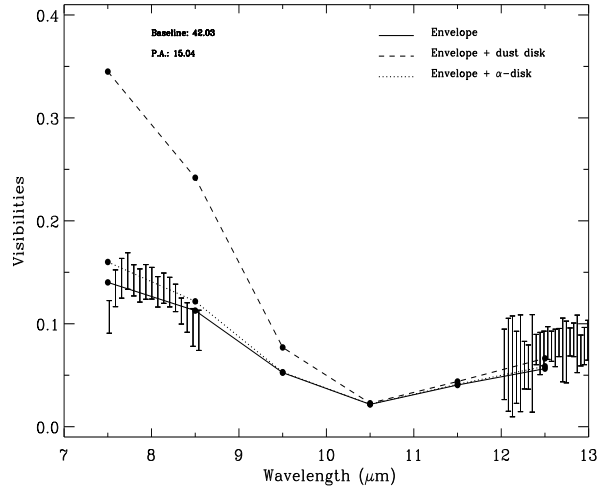}
  \caption[]{Model predictions for baseline configuration A. Represented are the preferred model (full line), with the
    addition of a dust disk ($M=0.25\,\Msun$, dashed line), and an optically thick disk interior to the dust sublimation 
radius (dotted line).}
  \label{frac}
\end{figure}

The resulting model visibilities of a protostellar envelope plus dust disk are
shown in Fig.\,\ref{frac}. It is clear that such a model violates the observed
MIDI visibilities shortward of the silicate absorption feature. At these
wavelengths the dust disk contributes $\sim 50\%$ to the total flux density,
and one can deduce that the disk has a visibility of approximately 70\%.  In
order for this model to be consistent with the observations, we compute that
the fractional contribution to the total 8\,\micron~flux by the dust disk
should be at most 1\%, if the disk emission at this wavelength is fully
unresolved. If it is marginally resolved (as it is) the 1\% upper limit
becomes smaller. By running models with increasingly smaller disk masses, we
find that this flux fraction is reached for a total mass of the dusty disk of
0.01\,\Msun, or equivalently $1.1\,10^{-7} (\alpha/0.01)\,\Msunyr$. The
SED of W33A is still fitted well by the dust disk$+$envelope model. We thus
find that the disk mass is very small compared to the envelope mass, and for
reasonable values of $\alpha$ that the dust disk accretion rate is more than
three orders of magnitude smaller than the envelope mass infall rate. These
results on the disk masses are in accordance with other studies of MYSOs (e.g. Murakawa et
al. 2008\nocite{2008A&A...490..673M}) and also with some evolved early-type
Herbig Be stars (Alonso-Albi et al. 2009\nocite{2009A&A...497..117A}). Would the disk
outer radius be much smaller than the assumed 500\,AU, and similar to the small outer radii 
found for the Herbig Be stars, then the disk mass should be scaled down accordingly; a 50\,AU 
outer radius results in a disk mass of $5\,10^{-4}\,\Msun$. A
small disk mass may imply that such a dust disk is not present, or that it is
simply extremely faint in the mid-IR due to the absence of gas and dust opacities 
in the model disk, which would lead to strong shielding from the stellar radiation field.


Most of the accretion luminosity however would be released within the 25\,AU
dust sublimation radius by the hot gaseous extension of the dust disk. 
Due to self-shielding, such a disk probably contains dust at smaller
radii than the 25\,AU sublimation radius determined from direct irradiation.
Inclusion of a self-luminous accretion disk within the dust
sublimation radius is required to fit the SED and the near-IR interferometric 
data of the Herbig B6e star \object{MWC\,147} (Kraus et al. 2008\nocite{2008ApJ...676..490K}).  Accretion
luminosity generated by an accretion rate of $7\,10^{-6}\,\Msunyr$ for this particular object 
is however not essential to model the star's mid-IR interferometry, which is already reasonably
well reproduced by a passive disk. We explore to what extent the mid-IR
interferometry of W33A is affected by the inclusion of an optically thick accretion disk.
We approximate the reslulting fluxes and visibilities with an optically thick,
geometrically thin $\alpha$ disk model (see Malbet \& Bertout
1995\nocite{1995A&AS..113..369M}; Malbet et al. 2007\nocite{2007A&A...464...43M})
and add these to the preferred envelope model in Table\,2. The disk is heated
by stationary accretion only and for an accretion rate equalling the mass
infall rate ($7.5\,10^{-4}\,\Msunyr$) the
generated accretion luminosity is $\sim 7\,10^{4}\,\Lsun$ for our adopted
central star. The spatial properties of the optically thick disk emission is such that 
99\% of the $N$-band is emitted within 50\,AU. Most of the
8\,\micron~flux is emitted at 20\,AU. The disk thus remains largely unresolved
on the employed VLTI baselines with MIDI. Disk accretion at a slower rate than the one
assumed are cooler and thus more compact. We choose the inner radius of the disk
to be at $1\,R_{*}$. The disk temperature falls below 1500\,K at 2.5\,AU, beyond
which the disk opacity is given by both gas and dust.

As in the previous paragraph, if
the preferred model visibilities are to be affected by the optically thick disk, the
ratio of disk emission to the emerging envelope emission should be at
least 1\% at 8\,\micron.  We calculate the resulting disk spectrum for the
said disk accretion rate and apply an extinction of $A_{\rm V}$=230 due to the
overlying envelope, as found in the line-of-sight to W33A from the preferred model. We show the resulting visibilities in
Fig.\,\ref{frac}. The figure shows that an $\alpha$ disk model embedded in a
protostellar envelope is compatible with the observed MIDI visibilities. The
model SED does not differ from the envelope only SED, provided that the total
luminosity of each model is chosen the same. The $\alpha$ disk contributes less than
1\% to the total 8\,\micron flux, despite its high luminosity. The envelope of W33A is simply
too opaque for any disk emission to be directly observed. MIDI observations of
MYSOs with less massive envelopes or at smaller inclinations however may well disclose the presence of an
accretion disk in such systems. The expected visibilities would be much larger
than the ones observed for W33A (depending on distance and accretion rate), considering the compactness
and brightness of the 8\,\micron~emission in accretion disks.

The popular YSO SED model grid developed by Robitaille et
al. (2006\nocite{2006ApJS..167..256R})\footnote{URL:
http://caravan.astro.wisc.edu/protostars/index.php} allows an estimate of the
extent to which MYSO SEDs are dominated by dust disk emission.  This YSO SED grid
makes use of the same 2-D axi-symmetric RT calculations as we use in this
paper (Whitney et al. 2003), and the disks in the SED model grid are therefore
dust disks. We extract all (135) SEDs corresponding to ZAMS hot stars with a
luminosity of $10^{3}<L/L_{\odot}<10^{6}$ and an envelope mass infall rate of
at least $\rm 10^{-4}\,M_{\odot}\,yr^{-1}$.  Inspection of the fractional disk
flux contribution at 8\,\micron~(just short of the extended wings of the
silicate absorption feature) shows that all except three of the extracted models
have a fractional contribution larger than 15\%, and half of the models have a
contribution larger than 70\%.  The geometry of the accretion disk determines
the total visibility at these wavelengths. The SED model grid thus contains a
preponderance of MYSO models with relatively high visibilities, similar to the
envelope plus dust disk model presented in Fig.\ref{frac}. At present, the MIDI
observations of MYSOs demonstrate the opposite trend however. Visibilities
presented for W33A are in line with the low $N$-band visibilities observed for
other MYSOs (see Linz et al. 2008\nocite{2008ASPC..387..132L}; Vehoff et al. 2008\nocite{2008ASPC..387..444V}; de
Wit et al. in prep.). The latter could equally well be explained in the
context of cavity wall emission as proposed here for W33A. If verified, then
MIDI has not (yet) revealed the presence of a MYSO accretion disk on size
scales of $\sim 100$\,AU, but has certainly the potential to do so.

\subsection{SED model grid fits} 
\begin{figure}
  \includegraphics[height=9cm,width=7cm,angle=90]{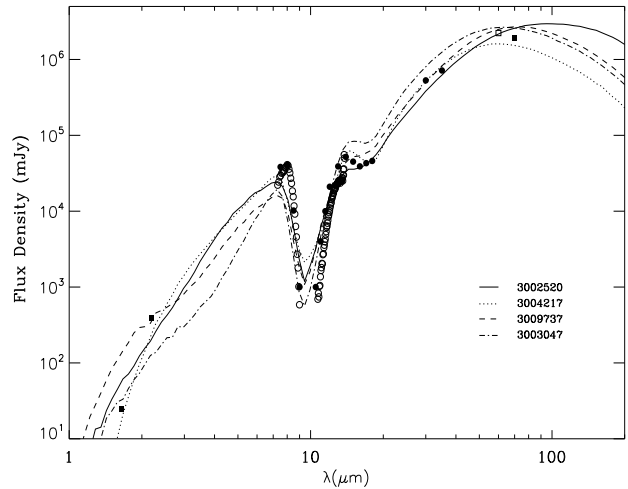}
  \caption[]{Predicted SEDs of the discussed models obtained  from the SED web fit
    procedure listed in Table\,\ref{webtab}.}
  \label{web}
\end{figure}

To what extent are models obtained from SED fits consistent with the spatial
information on milli arcsecond scales? To address this question, we apply the
web-based SED fitter procedure developed by Robitaille et
al. (2007\nocite{2007ApJS..169..328R}) to the W33A SED data shown in
Fig.\,\ref{web}. The procedure returns a ranking on the basis of a $\chi^{2}$
minimization method, and we discuss the grid models following this ranking.

\begin{figure}
  \includegraphics[height=9cm,width=7cm,angle=90]{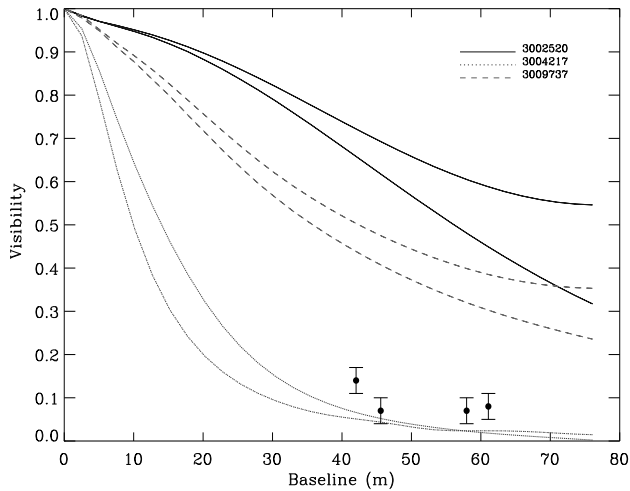}
  \caption[]{Predicted visibilities along the major (lower curve for each model) and
    minor axis (top curve for each model) at 8\,\micron~as a function of baseline for the discussed models 
    obtained from the SED web fit procedure (Table\,\ref{webtab}). MIDI visibilities for the
  four baselines are also shown.}
  \label{2520vis}
\end{figure}

The procedure delivers as best-fit a model consisting of a central hot star and a dust disk (model
grid number 3002520, see Table\,\ref{webtab}). We see from Fig.\,\ref{web}
that the grid model fit to the SED is acceptable, as expected. 
This particular grid model has a system inclination of $50\degr$, consistent
with the one of the reflection nebula. The best-fit model consists of a protostellar
envelope with a cavity opening angle relatively large compared to the one
derived from the scattering nebula. Moreover a fit of the model to the observed SED requires a distance of
1\,kpc, which is a factor $\sim 4$ too small. The discrepancy
between the modelled and observed SED is mainly at the depth of the silicate
feature. This depth is better accounted by dust partially consisting of ``warm silicates'' (see Ossenkopf et al. 1992)
as discussed previously; the appropriate set of optical constants is not used in the SED model grid. Broken down in its
various contributing components, we find that the grid model SED is completely
dominated by disk emission at wavelengths shortward of the silicate absorption
feature. The emission accounts for the relatively high flux levels observed 
shortward of the silicate feature. This is a common property of MYSO SEDs. Spherical and indeed
2D-axisymmetric RT codes for the protostellar envelope are unable to reproduce
these high flux levels (see e.g. de Wit et al. 2009\nocite{2009A&A...494..157D}).  Contrary to the good SED fit, Fig.\,\ref{2520vis}
shows that the corresponding visibilities are far too large. The
visibilities are obtained by producing images with the dust RT code using the
same parameter file that produced the SED. The grid model visibilities
are inconsistent simply because of the dominance of the (compact) disk
emission.  Clearly, a good model fit to the SED does not guarantee a proper
description of the object in consideration; the best-fit grid model violates
the spatial information delivered by MIDI.

\begin{figure}
  \includegraphics[height=9cm,width=9cm]{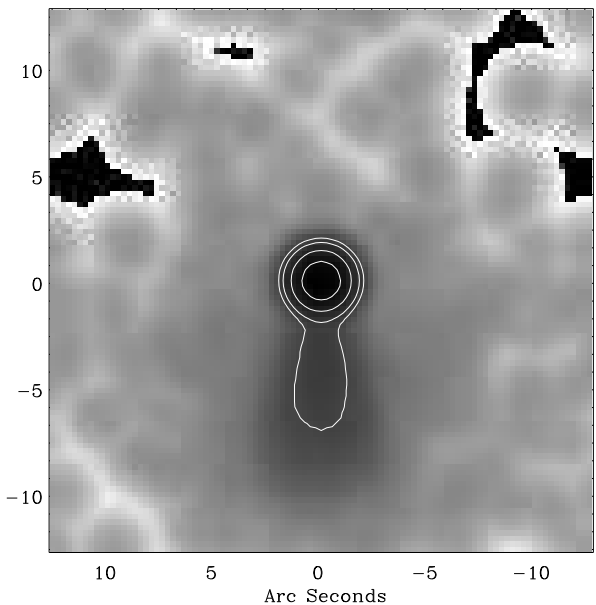}
  \caption[]{Logarithmically scaled $K$-band image for grid model 3004217 to be
    compared with the observations in Fig.\,\ref{scatmod}. The contours are
    at 1\%, 40\%, 50\% and 65\% of the peak flux. The image is convolved
with a 2-D Gaussian function with a FWHM of 1.6\arcsec.}
  \label{4217}
\end{figure}

If we use the known properties of W33A and limit the model distance to
be between 3.0 and 4.6\,kpc with a maximum foreground extinction of
$A_{v}=50$, then we find that the returned fitting models have a strong
preference for small inclinations. In the best 10 models, 7 have an
inclination of 18.19 degrees (the smallest present within the grid), also the best fit grid model (number 3004217). In
addition, the visibilities fit the MIDI observations reasonably well (Fig.\,\ref{2520vis}). The
model is characterized by a central cool star with a large stellar
radius. Such a description of the MYSO M8E-IR is preferred in Linz et
al. (2009\nocite{2009arXiv0907.0445L}).  In the case of W33A, the small inclination angles are 
inconsistent with the allowed range we derived from the
scattering nebula. The predicted $K$-band morphology of this model is shown in
Fig.\,\ref{4217}, and is strongly dominated by the central source 
which is in contradiction with the observations of W33A (Fig.\,\ref{scatmod}).
The ranked fourth model however has a more reasonable
inclination angle of $50\degr$ (number 3009737). The latter model consists of a hot
star and an accretion disk and encounters the previously discussed problems in
fitting the visibilities (Fig.\,\ref{2520vis}). If we would apply in selecting a model from the grid all the W33A
fiducial parameters ($40\degr<i<70\degr$, cavity opening angle
$10<2\theta<30$) for a hot star ($T_{\rm eff}>20\,000\,K$) without a
dust disk then the model complying with these constraints is ranked 245th
(model number 3003047). 

It is clear that the SED model grid can deliver models that 
fit the SED, but all the SED fitting models presented in
this section violate the MIDI visibilities. Regarding W33A, the SED grid
does not provide a fit to both SED and visibilities simply because the
appropriate models are not present in the grid (see also the discussion in 
Robitaille 2008\nocite{2008ASPC..387..290R}; Linz et al. 2009).

\begin{table*}
  {
    \begin{center}
      \caption[]{Models from the YSO SED model grid (Robitaille et
        al. 2006), obtained using the web-based fit procedure. The quoted distance is
required to fit the SED of W33A; $\theta$ is the outflow cavity opening angle; $A_{\rm V}$ is the applied foreground extinction.}
      \label{webtab} 
      \begin{tabular}{cccccccrcc}
        \hline
        \hline
        Model   & $i$       & $D$    & $T_{\rm eff}$ & $R_{*}$ &  $M_{\rm env}$ &$\dot{M}_{\rm infall}$  & $\theta$  & $M_{\rm disk}$ & $A_{\rm V}$\\  
                & ($\degr$) & (kpc)  & (kK)          &(\Rsun)  &  (\Msun)       &($\Msunyr$)            &  ($\degr$)& ($\Msun$)        &  \\
        \hline					     													      
        3002520 & 50        &  1.0   & $23.4$        & 4.5     & $7.5\,10^{2}$ & $1.1\,10^{-3}$          & 27.8      & 0.024          & 6.1  \\ 
        3004217 & 18        &  4.4   &  $4.8$        & 373     & $5.0\,10^{2}$ & $1.1\,10^{-3}$          & 2.7       & 0.0            & 9.2\\
        3009737 & 50        &  4.6   & $39.8$        & 7.0     & $1.3\,10^{3}$ & $3.8\,10^{-3}$          & 30.7      & 2.0            & 7.4\\ 
        3003047 & 41        &  3.0   & $34.4$        & 6.6     & $5.0\,10^{3}$ & $1.0\,10^{-3}$          & 10.7      & 0.0            & 0.0\\
       \hline
       \hline
      \end{tabular}
    \end{center}
  }
\end{table*}

\section{Conclusions}
\label{conclusions}
Observations with MIDI at the VLTI of the MYSO W33A indicate that $N$-band
emission on scales of 100\,AU is due to warm parts of the envelope close to 
the outflow cavity, i.e. the ``cavity walls''. The overlying envelope, when
described by a TSC type geometry
is characterised by an equivalent mass infall rate of order $10^{-3}\,\Msunyr$ and is
quite massive. We find that the presence of a dust disk located
beyond the dust sublimation radius violates the observed visibilities. In order to make the model
consistent with the observations the dust disk is required to be marginal in
mass. In contrast, the addition of an optically thick accretion disk
interior to the dust sublimation radius has little effect on the visibilities,
even when the disk dominates the luminosity. The disk emission is highly
extincted by the massive overlying protostellar envelope. MYSOs viewed at 
smaller inclination and/or with less massive protostellar envelopes have the
potential to reveal such accretion disks.

\begin{acknowledgements}
We would like to thank O. Chesneau for his prompt advices, and T. Fujiyoshi
for the W33A COMICS data. It is a pleasure to thank B. Whitney for discussions 
on the subject, and an anonymous referee for his clear remarks that improved 
this manuscript.
\end{acknowledgements}

\bibliographystyle{aa}

\end{document}